# Signal Recovery Using Splines

**Oleksandr Shumeyko, Ivan Devyatkin**
**Regional Communal Higher Educational Institution**
**"Strategy" the Institute for Entrepreneurship, Zhovti Vody, Ukraine**

**ABSTRACT:** Practically, for all real measuring devices the result of a measurement is a convolution of an input signal with a hardware function of a unit φ. We call a spline to be φ-interpolating if the convolution of an input signal with a hardware function of a unit φ coincides with the convolution of the spline with the hardware function.
In the following article we consider conditions imposed on the hardware function φ under which a second- and third-order φ-interpolating spline exists and is unique. Algorithms of φ-interpolating splines construction are written out.
**KEYWORDS:** Signal, spline, recovery, interpolation, asymptotically interpolation.

**Introduction**

The main concept of measurement theory is the concept of signal. Traditionally, the term 'signal' means disturbance of the physical environment characterizing the object of study, which can be registered some way. As a rule, during the measurement process the primary signal *f(t)* is converted into an electrical pulse which can be described with the ratio

$$\widetilde{f}(\tau) = \int_{-\infty}^{\infty} f(t)\varphi(t-\tau)dt \, ,$$

where $\widetilde{f}(\tau)$ is a registered output signal value of a primary converter and *φ* (*x*) is the transfer function of the primary converter that is the reaction function of the device to the disturbance appearing at the moment of signal registration.

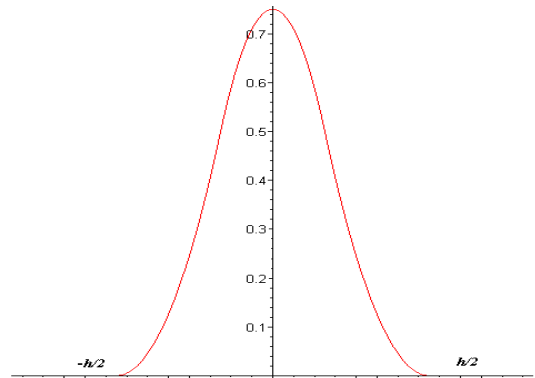

**Figure 1. Transfer function φ (t)**

**1 Problem definition**

For the discrete case, as a result of the registration according to the reference system $\{ih\}_{i \in Z}$ of a continuous signal $f(t)$ by the primary converter with the preset transfer function $\varphi$ $(t)$, we obtain a set of registered signal values

$$\tilde{f}_i = \frac{1}{h}\int_{-\infty}^{\infty} f(t)\varphi(t-ih)dt. \tag{1}$$

Thus, if we consider the problem of restoration of the signal $f(t)$ in terms of the set $\{\tilde{f}_i\}_{i \in Z}$ with the help of splines $s_r(f,t)$ of order $r$, then it is natural to satisfy the condition

$$\widetilde{s_r(f)}_i = \tilde{f}_i,$$

that is

$$\frac{1}{h}\int_{-\infty}^{\infty} s_r(f,t)\varphi(t-ih)dt = \frac{1}{h}\int_{-\infty}^{\infty} f(t)\varphi(t-ih)dt \tag{2}$$

216

From now on we will call the implementation of equality (2) as φ-interpolation.

As a rule, in scientific works dealing with recovery of functions and functionals with the help of splines, the usage of delta function $\delta(t)$ as a transfer function $\varphi(t)$ is considered. In this case $\tilde{f}_i = f_i = f(ih)$, and the problem reduces to building an traditional interpolating spline.

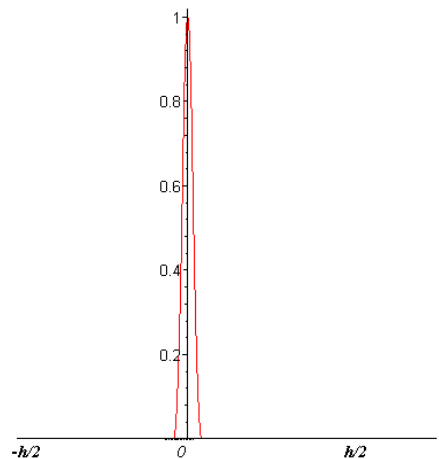

**Figure 2. Transfer function $\delta(t)$ (Dirac-function)**

Another known example of transfer function is the following step

$$\varphi(t) = \begin{cases} 1, & t \in \left[-\dfrac{h}{2}, \dfrac{h}{2}\right]; \\ 0, & t \notin \left[-\dfrac{h}{2}, \dfrac{h}{2}\right]. \end{cases}$$

In this case splines which satisfy the condition (2) are called interpolating on average or maintaining an average (see, for example, [LS97]).

In this article the problem of signal restoration using second- and third-order splines of minimum defect with uniform partition with a step *h* is considered.

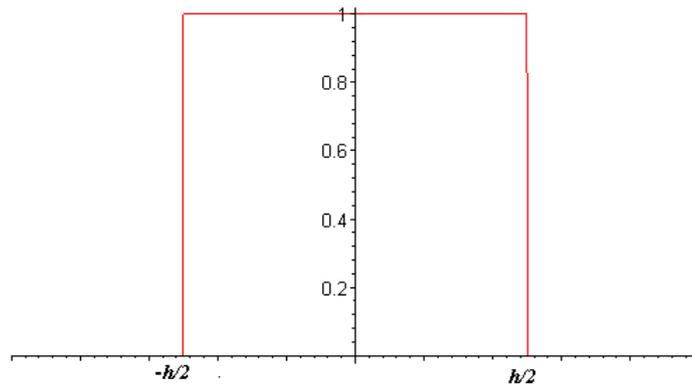

**Figure 3. Transfer function $\chi(t)$ (Heaviside step-function)**

Impose the following natural conditions on the transfer function

$$\varphi(x) \in C_{(-\infty,\infty)}, \varphi(t) = \varphi(-t), \varphi(t) = 0, t \notin [-h/2, h/2],$$
$$\frac{1}{h}\int_{-h/2}^{h/2}\varphi(t)dt = 1. \qquad (3)$$

Besides, let $m_k (k = 0,1,...)$ be the midpoint of order $k$ of the transfer function

$$m_k = \frac{1}{h^{k+1}}\int_{-h/2}^{h/2} t^k \varphi(t)dt,$$

and assume

$$m_k^+ = \frac{1}{h^{k+1}}\int_0^{h/2} t^k \varphi(t)dt,$$

and

$$m_k^- = \frac{1}{h^{k+1}}\int_{-h/2}^0 t^k \varphi(t)dt.$$

It is clear then that $m_{2k+1}^{+} = -m_{2k+1}^{-}$ and $m_{2k}^{+} = m_{2k}^{-}$ for k = 0, 1, 2, …, and also

$$m_{2k+1} = 0. \qquad (4)$$

**2 Main result**

At the beginning, let $r = 2$. It is known (see, for example, [Boo78]) that every parabolic spline of minimum defect with nodes at points $\{(i+1/2)h\}_{i \in Z}$ can be written as

$$s_2(t) = \sum_{i \in Z} c_i B_2(t - (i + \frac{1}{2})h), \qquad (5)$$

where $B_2(t)$ parabolic B-spline is written as

$$B_2(t) = \begin{cases} \frac{1}{2}(\frac{3}{2} + \frac{t}{h})^2, & t \in \left[-\frac{3}{2}h, -\frac{1}{2}h\right), \\ \frac{3}{4} - (\frac{t}{h})^2, & t \in \left[-\frac{1}{2}h, \frac{1}{2}h\right), \\ \frac{1}{2}(\frac{3}{2} - \frac{t}{h})^2, & t \in \left[\frac{1}{2}h, \frac{3}{2}h\right), \\ 0, & t \notin \left[-\frac{3}{2}h, \frac{3}{2}h\right), \end{cases}$$

Then for $t \in [-h/2, h/2]$

$$s_2(t) = c_{-1}\frac{1}{2}\left(\frac{1}{2} - \frac{t}{h}\right)^2 + c_0\left(\frac{3}{4} - \left(\frac{t}{h}\right)^2\right) + c_1\frac{1}{2}\left(\frac{1}{2} + \frac{t}{h}\right)^2 =$$

$$= \frac{1}{8}(c_{-1} + 6c_0 + c_1) + \frac{t}{2h}(c_1 - c_0) + \frac{t^2}{2h^2}(c_{-1} - 2c_0 + c_1),$$

consequently,

$$\widetilde{s_2(f)}_0 = \frac{1}{h}\int_{-h/2}^{h/2} s_2(t)\varphi(t)dt =$$

$$= \frac{1}{8}(c_{-1}+6c_0+c_1)\frac{1}{h}\int_{-h/2}^{h/2}\varphi(t)dt + \frac{c_1-c_0}{2h^2}\int_{-h/2}^{h/2} t\varphi(t)dt + \frac{c_{-1}-2c_0+c_1}{2h^3}\int_{-h/2}^{h/2} t^2\varphi(t)dt.$$

From the one above and from the data (3) and (4) we immediately derive

$$\widetilde{s_2(f)}_0 = \frac{c_{-1}+6c_0+c_1}{8} + \frac{c_{-1}-2c_0+c_1}{2}m_2.$$

Then the condition of φ-interpolation will become

$$\frac{c_{i-1}+6c_i+c_{i+1}}{8} + \frac{c_{i-1}-2c_i+c_{i+1}}{2}m_2 = \widetilde{f}_i,$$

or, which is the same

$$c_{i-1}\left(\frac{1}{8}+\frac{m_2}{2}\right) + c_i\left(\frac{3}{4}-m_2\right) + c_{i+1}\left(\frac{1}{8}+\frac{m_2}{2}\right) = \widetilde{f}_i \, (i \in Z). \tag{6}$$

To make the system (6) solvable, it is necessary to meet the condition of Hadamard's lemma (see, for example, [ZKM80]), that is the matrix of the system must have diagonal dominance, and since $m_2 > 0$, then this condition will become

$$\frac{3}{4}-m_2 - 2\left(\frac{1}{8}+\frac{m_2}{2}\right) = \frac{1}{2} - 2m_2 > 0.$$

Thus, if the transfer function of primary transfer $\varphi(t)$ is such that

$$0 < m_2 < \frac{1}{4}, \qquad (7)$$

then the φ-interpolating parabolic spline $s_2(f,t)$ exists and is unique.

Our further considerations follow the work [Lig81]. Let us choose

$$c_i^0 = \tilde{f}_i - \left(\frac{1}{8} + \frac{m_2}{2}\right)\Delta^2 \tilde{f}_i,$$

and let

$$s_{2,0}(f,t) = \sum_{i \in Z} c_i^o B_2\left(t - \left(i + \frac{1}{2}\right)h\right).$$

Then

$$\widetilde{s_{2,0}(f)}_i = \frac{c_{i-1}^0 + 6c_i^0 + c_{i+1}^0}{8} + \frac{c_{i-1}^0 - 2c_i^0 + c_{i+1}^0}{2} m_2 =$$

$$= c_i^0 + \frac{1}{8}\Delta^2 c_i^0 + \frac{1}{2}\Delta^2 c_i^0 m_2 = c_i^0 + \Delta^2 c_i^0 \left(\frac{1}{8} + \frac{1}{2}m_2\right) =$$

$$= \tilde{f}_i - \Delta^2 \tilde{f}_i\left(\frac{1}{8} + \frac{1}{2}m_2\right) + \Delta^2 \tilde{f}_i\left(\frac{1}{8} + \frac{1}{2}m_2\right) - \Delta^4 \tilde{f}_i\left(\frac{1}{8} + \frac{1}{2}m_2\right)^2 =$$

$$= \tilde{f}_i - \Delta^4 \tilde{f}_i\left(\frac{1}{8} + \frac{1}{2}m_2\right)^2,$$

That is, if $f \in C^4$, then a constant $\alpha_i$ exists, such that

$$\widetilde{s_{2,0}(f)}_i = \tilde{f}_i + \alpha_i h^4 f_i^{(4)} = \tilde{f}_i + O(h^4).$$

Following the idea of the work [4], let us choose

$$c_i^* = \sum_{n=o}^{\infty}(-1)^n\left(\frac{1}{8}+\frac{1}{2}m_2\right)^n \Delta^{2n}\widetilde{f}_i,$$

note that for convergence of series in the right member of the equality it is necessary and sufficient to satisfy the condition (7).

Assume

$$s_2^*(f,t) = \sum_{i\in Z} c_i^* B_2\left(t-\left(i+\frac{1}{2}\right)h\right), \qquad (8)$$

Then

$$\widetilde{s_2^*(f)}_i = c_i^* + \Delta^2 c_i^*\left(\frac{1}{8}+\frac{1}{2}m_2\right)^2 =$$

$$= \sum_{n=0}^{\infty}(-1)^n\left(\frac{1}{8}+\frac{1}{2}m_2\right)^n \Delta^{2n}\widetilde{f}_i + \sum_{n=0}^{\infty}(-1)^n\left(\frac{1}{8}+\frac{1}{2}m_2\right)^{n+1}\Delta^{2n+2}\widetilde{f}_i =$$

$$= (-1)^0\left(\frac{1}{8}+\frac{1}{2}m_2\right)^0 \Delta^0\widetilde{f}_i + \sum_{n=0}^{\infty}(-1)^n\left(\frac{1}{8}+\frac{1}{2}m_2\right)^n \Delta^{2n}\widetilde{f}_i +$$

$$+ \sum_{n=0}^{\infty}(-1)^n\left(\frac{1}{8}+\frac{1}{2}m_2\right)^{n+1}\Delta^{2n+2}\widetilde{f}_i =$$

$$= \widetilde{f}_i + \sum_{n=0}^{\infty}(-1)^{n+1}\left(\frac{1}{8}+\frac{1}{2}m_2\right)^{n+1}\Delta^{2n+2}\widetilde{f}_i + \sum_{n=0}^{\infty}(-1)^n\left(\frac{1}{8}+\frac{1}{2}m_2\right)^{n+1}\Delta^{2n+2}\widetilde{f}_i =$$

$$= \widetilde{f}_i.$$

Thus,

$$\widetilde{s_2^*(f)}_i = \widetilde{f}_i,$$

and therefore, the spline (8) is φ-interpolating.

Now, let $r = 3$. Any minimum defect cubic spline can be written as

$$s_3(t) = \sum_{i\in Z} c_i B_3(t-ih), \qquad (9)$$

where $B_3(t)$ cubic B-spline is

$$B_3(t) = \begin{cases} \dfrac{1}{2}\left(\dfrac{t}{h}\right)^3 - \left(\dfrac{t}{h}\right)^2 + \dfrac{2}{3}, & t \in [0,h), \\ \dfrac{1}{6}\left(2 - \dfrac{t}{h}\right)^3, & t \in [h,2h), \\ 0, & t \geq 2h. \end{cases}$$

and $B_3(-t) = B_3(t)$.

Then for $t \in [0,h]$

$$s_3(t) = \frac{1}{6}(c_{-1} + 4c_0 + c_1) + \frac{t}{2h}(c_1 - c_{-1}) + \frac{1}{2}\left(\frac{t}{h}\right)^2 (c_{-1} - 2c_0 + c_1) +$$

$$+ \frac{1}{6}\left(\frac{t}{h}\right)^3 (-c_{-1} + 3c_0 - 3c_1 + c_2)$$

and for $t \in [-h,0]$

$$s_3(t) = \frac{1}{6}(c_{-1} + 4c_0 + c_1) + \frac{t}{2h}(c_1 - c_{-1}) + \frac{1}{2}\left(\frac{t}{h}\right)^2 (c_{-1} - 2c_0 + c_1) -$$

$$- \frac{1}{6}\left(\frac{t}{h}\right)^3 (-c_1 + 3c_0 - 3c_{-1} + c_{-2}).$$

From that we immediately derive

$$\int_{-h/2}^{h/2} \varphi(t)s_3(t)dt = \frac{1}{6}(c_{-1} + 4c_0 + c_1) + \frac{1}{2}(c_1 - 2c_0 + c_{-1})m_2 +$$

$$+ \frac{1}{6}(-c_{-1} + 3c_0 - 3c_1 + c_2)m_3^+ - \frac{1}{6}(-c_1 + 3c_0 - 3c_{-1} + c_{-2})m_3^-.$$

223

Noting that $m_3^- = -m_3^+$, we have

$$\tilde{\mathfrak{s}}_{30} = \frac{1}{6}(c_{-1} + 4c_0 + c_1) + \frac{1}{2}(c_1 - 2c_0 + c_{-1})m_2 +$$
$$+ \frac{1}{6}(c_{-2} - 4c_{-1} + 6c_0 - 4c_1 + c_2)m_3^+ =$$
$$= c_0 + \frac{1}{6}\Delta^2 c_0 \left(\frac{1}{6} + \frac{m_2}{2}\right) + \Delta^4 c_0 \frac{m_3^+}{6}.$$

The condition of φ-interpolation in this case becomes

$$\frac{m_3^+}{6}c_{i-2} + c_{i-1}\left(\frac{1}{6} + \frac{m_2}{2} - 2\frac{m_3^+}{3}\right) + c_i\left(\frac{2}{3} - m_2 + m_3^+\right) +$$
$$+ c_{i+1}\left(\frac{1}{6} + \frac{m_2}{2} - 2\frac{m_3^+}{3}\right) + \frac{m_3^+}{6}c_{i+2} = \tilde{f}_i (i \in Z).$$

Then for a φ-interpolating cubic spline to exist and be unique, it is necessary and sufficient to satisfy the condition

$$\left|\frac{2}{3} - m_2 + m_3^+\right| - 2\left|\frac{1}{6} + \frac{m_2}{2} - 2\frac{m_3^+}{3}\right| - \frac{m_3^+}{3} > 0.$$

The solution will be the set $D = D_1 \cup D_2$ where

$$D_1 = \left\{(m_3^+, m_2): 0 < m_2 < \frac{5}{3}, 0 < m_3^+ < \frac{3}{2}, m_3^+ - m_2 + \frac{1}{6} > 0, \frac{4}{3}m_3^+ - m_2 < \frac{1}{3}\right\}$$

$$D_2 = \left\{(m_3^+, m_2): 0 < m_2 < \frac{5}{3}, \frac{1}{4} < m_3^+ < \frac{3}{2}, \frac{4}{3}m_3^+ - m_2 > \frac{1}{3}\right\}$$

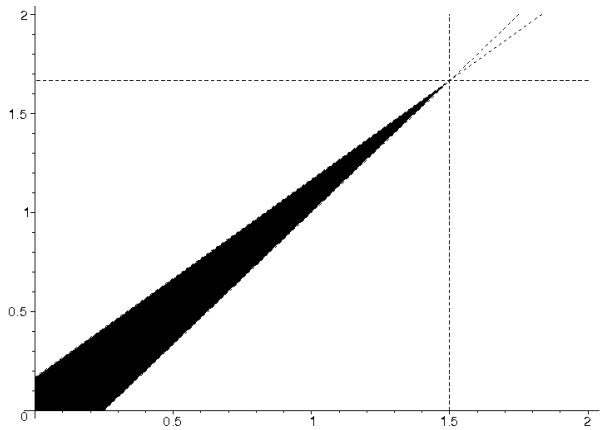

**Figure 4. Set $D_1$**

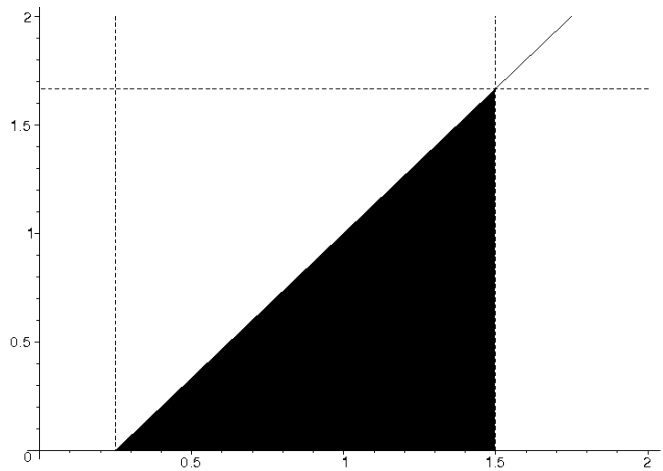

**Figure 5. Set $D_2$**

Same as above, let us choose

$$c_i^0 = \tilde{f}_i - \frac{1}{6}\Delta^2 \tilde{f}_i\left(\frac{1}{6} + \frac{m_2}{2}\right) - \Delta^4 \tilde{f}_i \frac{m_3^+}{6}$$

and let

$$s_{3,0}(f.t) = \sum_{i \in Z} c_i^0 B_3(t - ih).$$

225

Then

$$\widetilde{s_{2,0}(f)}_i = c_i^0 + \frac{1}{6}\Delta^2 c_i^0 \left(\frac{1}{6} + \frac{m_2}{2}\right) + \Delta^4 c_i^0 \frac{m_3^+}{6} =$$

$$= \widetilde{f}_i - \frac{1}{6}\Delta^2 \widetilde{f}_i \left(\frac{1}{6} + \frac{m_2}{2}\right) - \Delta^4 \widetilde{f}_i \frac{m_3^+}{6} +$$

$$+ \frac{1}{6}\Delta^2 \left(\widetilde{f}_i - \frac{1}{6}\Delta^2 \widetilde{f}_i \left(\frac{1}{6} + \frac{m_2}{2}\right) - \Delta^4 \widetilde{f}_i \frac{m_3^+}{6}\right)\left(\frac{1}{6} + \frac{m_2}{2}\right) +$$

$$+ \Delta^4 \left(\widetilde{f}_i - \frac{1}{6}\Delta^2 \widetilde{f}_i \left(\frac{1}{6} + \frac{m_2}{2}\right) - \Delta^4 \widetilde{f}_i \frac{m_3^+}{6}\right)\frac{m_3^+}{6} =$$

$$= \widetilde{f}_i - \frac{1}{6^2}\Delta^4 \widetilde{f}_i \left(\frac{1}{6} + \frac{m_2}{2}\right)^2 - \Delta^6 \widetilde{f}_i \left(\frac{1}{6} + \frac{m_2}{2}\right)^2 \frac{m_3^+}{18} - \Delta^8 \widetilde{f}_i \left(\frac{m_3^+}{6}\right)^2.$$

that is, if $f \in C^4$, then a constant $\beta_i$ exists, such that

$$\widetilde{s_{3,0}(f)}_i = \widetilde{f}_i + \beta_i h^4 f_i^{(4)} = \widetilde{f}_i + O(h^4),$$

that is, spline (10) is asymptotically ('almost') φ-interpolating.

**References**

[Boo78] **C. de Boor.** - *A Practical Guide to Splines*, Springer-Verlag, 1978

[LS97] **A. Ligun, A. Shumeyko** - *Asymptotical method recovery's the curves* [in Russian], Kyiv, Math. Inst. NAN Ukraine, 1997, 358 p.

[Lig81] **A. Ligun** - *Approximation of differentiable functions by local splines of minimal defect*, Ukr.math.journ., 1981, v.33, №5, p. 691-693

[ZKM80] **Y. S. Zavyalov, B. I. Kvasov, V. L. Miroshnichenko** - *Method of Spline-Functions* [in Russian], Moscow, 1980